\documentclass[aps,prl,twocolumn,superscriptaddress,floatfix,amssymb]{revtex4-1}

\usepackage{graphicx}
\usepackage[dvipsnames]{xcolor}
\begin{document}

\title{Persistently Non-Gaussian Metastable Liquids}

\author{Vinay Vaibhav}
\email{vinay.vaibhav@uni-goettingen.de}
\affiliation{Institut für Theoretische Physik, University of Göttingen,
Friedrich-Hund-Platz 1, 37077 Göttingen, Germany}

\author{Tamoghna Das}
\email{tamoghna@staff.kanazawa-u.ac.jp}
\affiliation{WPI Nano Life Science Institute (WPI-NanoLSI), Kanazawa University, 
Kakuma-machi, Kanazawa, 920-1192 Japan}

\author{Suman Dutta}
\email{d\_suman1@cb.amrita.edu (Corresponding Author)}
\affiliation{Amrita School of Artificial Intelligence, Coimbatore, Amrita Vishwa Vidyapeetham, India}

\begin{abstract}
   Particles undergoing Fickian diffusion within smooth energy landscapes exhibit Gaussian statistics. However, this Gaussian behavior is often elusive in complex liquids, where particle dynamics within spontaneously fluctuating or spatio-temporally heterogeneous environments lead to a breakdown of ergodicity and time-reversal symmetry. This is usually caused by extreme particle movements or sudden dynamical arrest. Such situations are prevalent in dense metastable liquids exhibiting slow flow or {cooperative movements, facilitated by} cage-breaking. We investigate the dynamics of glassy systems driven by either thermal, external, or environmental fluctuations. Despite their differences, our findings reveal that particle motion is universally affected by large deviations, resulting in non-Gaussian tails persisting over multiple decades in time. We further discuss the underlying dynamical aspects.
\end{abstract}                   

\maketitle

\section{1. Introduction}

The gentle tapping of a sand plate and the rapid cooling of molten glass—though seemingly unrelated—both exemplify the physics of glassy systems: far from equilibrium and trapped in locally metastable states. \cite{berthier2011theoretical, biroli2001metastable}. This perspective provides a unifying notion for understanding the onset of flow, failure and organization in a diverse range of physical systems \cite{nicolas2018deformation, ozawa2018random, berthier2025yielding, frey2005brownian, schuh2007mechanical, bera2020quantitative, vaibhav2023controlled, di2024brownian}. It includes sharply increasing viscosity in molecular and metallic glasses \cite{berthier2011theoretical}, shear thinning and thickening in glasses and suspensions \cite{singh2025viscosity, vaibhav2022rheological, morris2020shear}, counterintuitive flow properties of granular materials \cite{rao2008introduction} and synaptic plastic activity \cite{khona2022attractor}. Recent theoretical and experimental advances have further connected these observations from soft condensed matter to interdisciplinary sciences, such as flocking birds \cite{cavagna2014bird}, dense epithelial monolayers \cite{angelini2011glass, czajkowski2019glassy}, developing tissues \cite{kim2021embryonic}, financial markets \cite{sinha2010econophysics} and intelligent robotic swarms \cite{rubenstein2014programmable}. Central to this theme is the principle that the statistics of the underlying constituent variables in these non-equilibrium systems universally deviate from Gaussianity \cite{chaudhuri2007universal, wang2012brownian, chubynsky2014diffusing, jain2017diffusing, cherstvy2019non, barkai2020packets, dandekar2020non,  vaibhav2024entropic, chandrasekhar1943stochastic, wang2012brownian, shell2016coarse, cherstvy2019non}. Therefore, developing a common pedagogical connection for fundamental understanding, is significant.

A key manifestation of these non-equilibrium behaviors is heterogeneous dynamics, which has emerged as a unifying phenomenology across diverse interacting many-body systems \cite{kegel2000direct, martens2011connecting, garrahan2011dynamic, berthier2011dynamical, wang2012brownian, starr2013relationship, jung2023predicting, ozawa2023elasticity, vaibhav2024entropic}. Recent scientific evidences reveal that the dynamically distinct population classes in glassy systems coexist by well-separated mobilities or relaxation timescales \cite{kegel2000direct, gao2007direct, andersen2005molecular, martens2011connecting, jung2023predicting, vaibhav2024entropic, Dutta2019-pg,das2018quantifying, das2019three}. In such systems, spatial regions relax in correlated yet disparate ways, resulting in pronounced differences in local relaxation times and transport properties \cite{szamel2006time, karmakar2014growing}. This has been an intuitive theme of a wide variety of amorphous systems that have been proposed recently \cite{binder2011glassy,dutta2025activity, dutta2023creep, vaibhav2024entropic}.

In these systems heterogeneity in dynamics is ubiquitous. The theoretical basis for such heterogeneity can be generalized using the phenomenology of \textit{Brownian yet non-Gaussian diffusion} \cite{chubynsky2014diffusing, wang2012brownian}, wherein complex particle dynamics are ascribed to \textit{diffusing-diffusivity}, unlike the Fickian theory that considers a unique diffusion for typical particle dynamics. This can be further understood using the notion of potential energy landscape, that considers the characteristics of the lowest-energy {\em inherent} structures within a metastable liquid \cite{heuer2008exploring, berthier2011dynamical}. In this multi-dimensional hyperspace with intricate topology, the equilibrium liquid is represented by stable minima \cite{sastry1998signatures}, while activated states occupy higher energy levels. Once a system is pushed out of this stable zone, it has to cascade through numerous metastable states on its way, in search of the global minimum \cite{binder2011glassy, sastry1998signatures}. However, the intricate and largely unknown topography of the landscape impedes this relaxation pathway, causing the system to become locally trapped within a meta-basin for a time longer than the experimentally accessible scale \cite{biroli2001metastable}. Heterogeneous dynamics at the collective scale naturally arises, as part of the system traverses this rugged basin topography, with some particles rapidly descending along local potential gradients while others slowly creep to the nearest local energy barriers \cite{mandal2021study, maloney2006amorphous}.

At the individual particle scale, heterogeneity manifests itself as the local phenomenon of caging \cite{wang2009anomalous, wang2012brownian, chubynsky2014diffusing, binder2011glassy}. Here, dynamic crowding transiently restricts the motion of a given particle by its neighbors, effectively forming a temporary spatial trap \cite{wang2012brownian, binder2011glassy, vaibhav2022finite}. Escape from this spatiotemporal confinement is facilitated by fluctuations, whether of thermal or other (e.g., athermal) origins. Such caged dynamics contrasts sharply with the conventional diffusive motion in simple liquid at equilibrium \cite{binder2011glassy, einstein1905motion}. Consequently, the temporal evolution of self-displacement fluctuations serves as an excellent indicator of the system's state, distinguishing between the equilibrium and metastable states. While for the ideal Brownian motion, the displacement distributions are Gaussian \cite{einstein1905motion}, any deviation from normality acts as a precursor of metastability or vulnerability. This is common in coarse-grained models of soft condensed matter, that help explain complex biological and cognitive dynamics, often exhibiting collective intelligence, arising from inhibition, excitation, self-regulation, and synchronization among the interacting units \cite{garrahan2011dynamic, berthier2011dynamical, Lowen2025-if}. Investigating such emergent behavior, thus, requires foundation that go beyond typical Gaussian assumptions, motivating the use of information theory and superstatistical approaches to account for rare events and fluctuating local environments \cite{dandekar2020non, vaibhav2024entropic}.

Here, we investigate the spatio-temporally heterogenous dynamics in different amorphous systems, exposed to a variety of conditions, interact via different force fields, when they are influenced by thermal or athermal fluctuations. We analyze single particle dynamics for extremely dense liquids, using displacement statistics obtained from the molecular dynamics simulations, done for the passive, stressed or active glassy systems. We show that particle dynamics is generically affected by the heterogeneity of the environment or applied perturbation under these situations, leading to persistent non-Gaussian displacement distributions. We discuss its distinctness from Fickian diffusion where such distributions are Gaussian with its standard deviation grows with elapsed time, at a fixed temperature. 

\section{2. Model and Methods} 
We consider three  different models of amorphous glasses under diverse physical conditions: (a) a model supercooled liquid, (b) a soft amorphous solid approaching its yielding transition, and (c) a model of flowing dense active matter. These models have previously demonstrated considerable success in capturing key phenomenological aspects of their respective physical class \cite{vaibhav2024entropic, liu2021elastoplastic, mandal2020extreme}.  \textbf{}
\subsection{2.1 Model}
We consider interacting particulate system where the dynamics of the $i$-th particle is expressed as{,}
\begin{equation}
   m_i\ddot{{\bf r}}_i= \mathsf{F}_i + \mathsf{B}_i,
\end{equation}
where $\mathsf{F}_i$ is a generalized force that contains all {\em systematic} forces experienced by a body at position ${\bf r_{i}}$ through interactions and/or otherwise, $\mathsf{F}={\bf F}^C+{\bf F}^D$ where ${\bf F}^C=-\partial_rU$ is a conservative force resulting from the pair-wise interaction potential $U$. ${\bf F}^D$ acts as a placeholder for all other possible forces of dissipative nature. The {\em random} forces of thermal origin $\zeta^T$ and of athermal nature (external perturbation and/or internal activity) $\zeta^A$, experienced by each body, are contained within $\mathsf{B}=\zeta^T+\zeta^A$. The random forces are delta-correlated in both space and time unless stated otherwise. The pair-wise interaction between the two components $(\alpha, \beta \in \{A, B\})$ is modeled by the well-known Lennard-Jones potential \cite{allen2017computer}:
\begin{equation}
U =4\epsilon^{\alpha\beta}\left[ \left(\frac{\sigma^{\alpha\beta}}{r}\right)^{12}-\left(\frac{\sigma^{\alpha\beta}}{r}\right)^6\right].
\end{equation}
Setting the energy- and length-scales by $\epsilon$ and $\sigma$, we perform molecular dynamics simulation to generate the trajectories of the systems for several decades of time unit, $\tau=\sqrt{\sigma^2/\epsilon}$. As these scales crucially depend on our choices of parameters for each particle type, such details along with other relevant thermodynamic information are tabulated in table~\ref{table1}. For all cases, the mass of individual bodies is set to unity, and thus dropped from the equation. Below we introduce our model systems briefly. This is followed by a discussion about our observables and how they have been computed from the simulation data.

\begin{table*}
\begin{tabular}{|p{1.5cm}|p{1.0cm}|p{2.0cm}|p{1.7cm}|p{1.2cm}|p{1.8cm}|p{1.0cm}|p{1.0cm}|p{1.0cm}|p{2.5cm}|}
 \hline
 \multicolumn{10}{|c|}{Simulation Models} \\
 \hline
 Model & $\rho$ & $N_{A}:N_{B}$ & $\sigma_{AA}$ & $\sigma_{AB}$ & $\sigma_{BB}$ & $\epsilon_{AA}$ & $\epsilon_{AB}$ & $\epsilon_{BB}$ &Remarks\\
 \hline
 Model 1   & 1.20    &80:20&   1.0& 0.80 &0.88 &1.0 &1.5&0.5&Thermal (3D)\\
 Model 2 &   1.02  & $(1+\sqrt{5}):4$  & $2\sin(\pi/5)$ & 1.00 & $2 \sin(\pi/10)$ &0.5 &1.0 &0.5&Athermal (2D)\\
 Model 3 & 1.20 & 65:35&  1.0& 0.80 &0.88 &1.0 &1.5&0.5&Athermal (2D)\\
 \hline
\end{tabular}
\caption{We use these models for a range of system sizes,  $N=1000-102400$ in our simulations in two or three dimensions (2D or 3D). For each cases, we generate $16-32$ trajectories with independent origins and statistics was gathered over these trajectories. Also we check for a range of respective controlling parameters.}
\label{table1}
\end{table*}

{\bf Model 1 - A model supercooled liquid:} The first system that we simulate is a model glass-former, in thermal equilibrium without and drive or dissipation. The model, in particular, is widely known as the Kob-Anderson 80:20 model \cite{kob1995testing}. It consists of a binary mixture of particles at a constant density that interact via LJ potential in three dimensions, kept at a constant temperature $T$. We use molecular dynamics simulations to equilibrate the system at these temperatures . We observe particle dynamics within a range of temperatures in the supercooled regime, above its glass-transition. More details about the model and simulation can be found in Refs. \cite{vaibhav2020response, vaibhav2024entropic}.
    
{\bf Model 2 - Driven amorphous soft solid:} We investigate another popular model of amorphous system \cite{lanccon1986thermodynamical} to address the microscopic onset of flow in driven glassy matter in two dimensions (2D), motivated by the material behavior of the yield stress fluids in athermal conditions without any active degree of freedom. Here, particle dynamics is modeled via the dissipative particle dynamics (DPD), where the dissipative force, ${\bf F}^D=-\xi w^{2}(r) ~ (\hat{{\bf r}} ~. ~{\bf v})\hat{{\bf r}}$, with $\xi$ as the friction coefficient and $w(r)=\sqrt{(1-r/r_{c})}$ as the weight function controlled by the ratio of pair separation to cutoff distance \cite{soddemann2003dissipative}. The framework uses a thermostat by canceling the drifts originated due to shear. The imposition of a constant macroscopic shear-stress, $\Sigma$ on the system was implemented using a feedback scheme, satisfying a macroscopic equation of motion, ${d\gamma}/{dt}\propto (\Sigma-\Sigma_{M})$ where $\Sigma_{M}$ is the Irving-Kirkwood stress. Detailed information on this model and preparation protocols can be found in Refs. \cite{cabriolu2019precursors, liu2021elastoplastic}.
    
{\bf Model 3 - An active system under extreme self-propulsion:} This model considers dense fluids with intrinsic self{-}propulsion that is not allowed to self-evolve with time, motivated by the infinitely persistent behavior of the cells \cite{angelini2011glass, bi2016motility}. The state in the complete absence of thermal fluctuations can exist in solid or fluid state, depending on the magnitude of self-propulsion forces at particle scale, that alone competes with microscopic interactions. The particles at the coarse-grained scale experience the persistent propulsion forces, $\zeta^{A}_i = f\hat{\eta_{i}}$ and a damping due to inertia. Here, $f$ is the magnitue of the active propulsion forces and  $\hat{\eta} \equiv (\sin{\theta}, \cos{\theta})$ assigns the direction of the random self-propulsion forces, with $<\zeta^{A}>(=f\sum_{i=1}^{N}\hat{\eta_{i}}) \rightarrow 0$ ensures that the entire system is not constantly driven as a whole, so the dynamics is only influenced by the intrinsic fluctuations. More details about this model can be found here \cite{mandal2020extreme, dutta2025activity}.

Model 1 simulates a glassy liquid at thermal equilibrium, while Model 2 shows onset of flow only beyond a critical threshold in imposed shear-stress, even in the complete absence of thermal fluctuations. Using the third model, we simulate situations of an active athermal {dense flow}. We simulate these models using LAMMPS \cite{lammps} and gather displacement statistics from the trajectories, at a single particle level.

For Model 1, we ensure equilibration in all the trajectories, starting from independent initial configurations. We average the observables over the equilibrium configurations, also over trajectories. The Model 2 has transient dynamics, where such averaging is not possible, unless the system reaches steady states from the initially jammed state. For Model 3, we again ensure that the system reaches steady state after a long run. For both Model 1 and 2, we only average over the trajectories for the improved averages. 

\begin{figure*}[t]
    \centering
    \includegraphics[width=0.98 \linewidth]{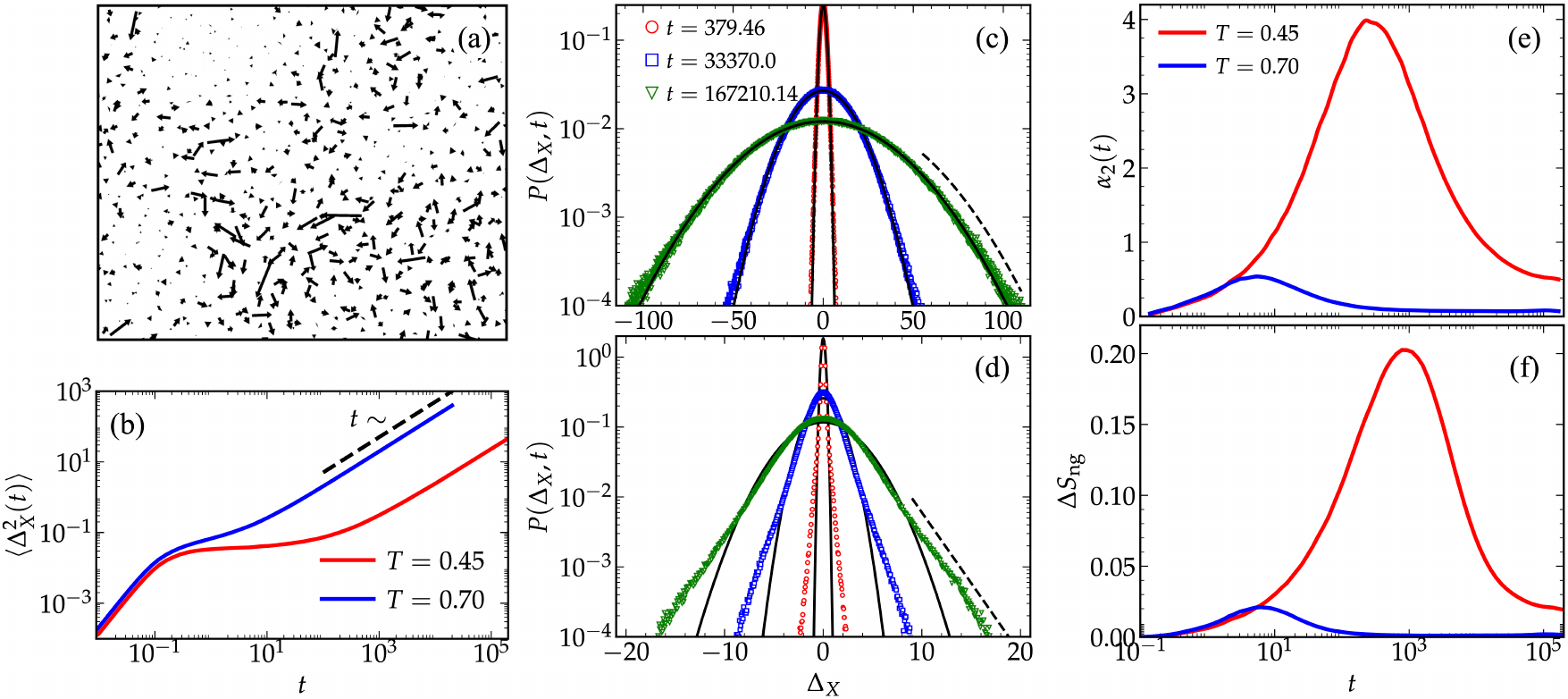}
    \caption{Dynamic heterogeneity in supercooled liquid: (a) Spatial map of particle displacements indicating heterogeneous dynamics at $T= 0.45$. Length of the arrows represents the magnitude of the displacements at single particle level in a 2D slice of the three dimensional system. (b) Mean-squared displacement as a function of time for two different temperatures as indicated. At low temperature, prominent plateau can be seen beyond the ballistic regime. At long times, the mean squared displacements grow linearly in time (shown by dotted line), although they may not be indicator of onset of Fickian-diffusion. (c,d) Distribution of particle displacements (shown with open symbols) and the corresponding nearest Gaussian constructions (shown with solid black lines) at three different time intervals as indicated; for (c) low temperature ($T=0.45$) and (d) high temperature ($T=0.70$). Both the particle distributions and its nearest Gaussian has identical first and second moments. (e,f) Measurements of Non-Gaussianity using conventional and optimal methods: conventional non-Gaussian parameter $\alpha_2$ (obtained from the ratio of the kurtosis to square of the second moment) and negentropy $\Delta S_{\rm ng}$ (obtained from the KL-divergence between the displacement distributions and its nearest Gaussian, as shown in figs. (c-d)) as a function of time for the system at temperatures $T = 0.45, 0.70$. Here, we used Model 1 to simulate these data in three dimensions for a system size $N=1000$.}
    \label{fig1}
\end{figure*}

\subsection{2.2 Dynamical Quantities}
We use the simulated trajectories to analyse the particle dynamics. In two dimensions, or in its projection, from the position of the $i$-th particle at time $t$, $\mathbf{r}_i(t) = (x_i(t), y_i(t))$, we calculate the following quantities to characterize the heterogeneity in dynamics:

{\bf Spatial displacement:} The spatial displacement provides a visual representation of displacement vector over a specific time interval $t$. For each particle $i$, its displacement vector starting from a reference time $t_0$ is calculated as:

\begin{equation}
    \Delta_\mathbf{r}^i(t_0, t) = \mathbf{r}_i(t_0 + t) - \mathbf{r}_i(t_0)
    \label{eq:disp_map}
\end{equation}

$\Delta_\mathbf{r}^i$ are typically plotted as arrows originating from the reference positions $\mathbf{r}_i(t_0)$, whose length is given by the magnitude of this vector, while the direction is given by the angle it creates with $x-$axis, for an interval $t$.

{\bf Mean squared displacement (MSD):} It quantifies the average squared to distance traveled by the particles within an interval $t$, from the reference origin $t_{0}$. We average over possible time origins $t_0$ available from a trajectory :

\begin{equation}
    <\Delta_r^{2}(t)> = \langle |\mathbf{r}_i(t_0 + t) - \mathbf{r}_i(t_0)|^2 \rangle_{i, t_0, Traj}
    \label{eq:msd}
\end{equation}

where $\langle \dots \rangle_{i, t_0, Traj}$ denotes an average over all particles $i=1, \dots, N$, available time origins $t_0$ and the trajectories. For isotropic systems, we calculate this in each direction and an average over such possibilities has also been considered. 

{\bf Displacement distributions:} The displacement distributions (also known as the self-van Hove function \cite{van1954correlations}), $P(\Delta_r, t)$, describe the probability distribution function of particle displacements ($\Delta_r=(\Delta_x,\Delta_y)$) within the time interval $t$:
\begin{equation}
    P(\Delta_r, t) = \left\langle \frac{1}{N} \sum_{i=1}^{N} \delta \left( r - |\mathbf{r}_i(t_0 + t) - \mathbf{r}_i(t_0)| \right) \right\rangle_{Traj}
    \label{eq:disp_dist}
\end{equation}

{The steady state displacement distribution was computed by performing ensemble averages over independent samples and temporal averages over multiple time origins.} 

{\bf Non-Gaussian parameter:} The non-Gaussian parameter, often denoted as $\alpha_2(t)$, quantifies the deviation of the particle displacement distribution ( as $P(\Delta_X, t)$) from a Gaussian distribution. For the one directional projection, it is defined using the second and fourth moments of the displacement \cite{kob1995testing}:

\begin{equation}
    \alpha_2(t) = \frac{\langle |\Delta_\mathbf{x}(t)|^4 \rangle}{3 \left( \langle |\Delta_\mathbf{x}(t)|^2 \rangle \right)^2} - 1.
    \label{eq:alpha2}
\end{equation}

A value of $\alpha_2(t) = 0$ indicates purely Gaussian dynamics for the interval. Positive values indicate the possibility of {\it fat tails} in the displacement distribution or the presence of extreme statistics, characteristic of heterogeneous dynamics.

{\bf Negentropy:} Complementary to moment-based measures, we employ the non-Gaussian information, $\Delta S_{ng}(t)$ \cite{dandekar2020non, ivan2012measure, shell2016coarse}, an information-theoretic metric based on Kullback-Leibler (KL) divergence. It quantifies the statistical distance metric between the observed particle displacement distribution $P(\Delta_\mathbf{x},t)$ and its {\it nearest Gaussian} counterpart $P^G_S(\Delta_\mathbf{x},t)$, which is a Gaussian distribution sharing the same mean $\mu(t)$ and variance $\sigma^2(t)$ as $P(\Delta_\mathbf{x},t)$. $\Delta S_{ng}(t)$ is defined as:
\begin{equation}
     \Delta S_{ng}(t) = A~ D_{KL} (P (\Delta_\mathbf{x},t)|| P^{G}_{S} (\Delta_\mathbf{x},t) )
     \label{eq:defng_condensed}
\end{equation}
where $A=1$ in this study and $D_{KL}(P||Q)=\int P(x)\log[P(x)/Q(x)] dx$ is the KL divergence \cite{kullback1951information}. $D_{KL}$ is non-negative and zero if and only if $P(x)=Q(x)$. Thus, $\Delta S_{ng}(t)$ measures the information lost when approximating $P(\Delta_\mathbf{x},t)$ with $P^G_S(\Delta_\mathbf{x},t)$, vanishing only if $P(\Delta_\mathbf{x},t)$ is itself Gaussian.

While the standard non-Gaussian parameter $\alpha_2(t)$ primarily reflects kurtosis, $\Delta S_{ng}(t)$ offers a more comprehensive measure of the overall statistical difference between the displacement distribution and its nearest Gaussian approximation \cite{vaibhav2024entropic}.

{\bf Local MSD:} The local MSDs are computed by coarse-graining the mean squared displacements at spatial scales. The system is divided into a grid of cells. For each cell $j$, at a site $(x,y)$, a local MSD, denoted by $(\Delta_r^{2})_j(x,y;t)$, is calculated by averaging the squared displacements only of those particles $i$ whose initial positions $\mathbf{r}_i(t_0)$ fall within that cell $j$ at the reference time $t_0$:
\begin{equation}
    (\Delta_r^{2})_j(x,y;t) = \frac{1}{N_j(t_0)} \sum_{i \, | \, \mathbf{r}_i(t_0) \in \text{cell } j} |\mathbf{r}_i(t_0 + t) - \mathbf{r}_i(t_0)|^2
    \label{eq:local_msd}
\end{equation}
where $N_j(t_0)$ is the number of particles whose initial position at $t_0$ is within cell $j$ locted at $(x,y)$. This calculation has been performed for a specific $t_0$ to get an instantaneous map, to obtain a time-averaged local MSD map.

{\bf Time series of displacements:} It contains the incremental displacements of individual particles over consecutive, typically short, time steps. For a single particle $i$, we calculate its displacement magnitude (or components) for a small interval $t$:

\begin{equation}
    \Delta_{r}^i(t_0) = \mathbf{r}_i(t_0 + t) - \mathbf{r}_i(t)
    \label{eq:disp_series}
\end{equation}
In addition, the x and y components of these incremental displacements were also tracked.

\section{3. Results} 
{In this section, we discuss results obtained from the analysis of the models described in section 2.1. These models have been investigated under a range of conditions chosen to model complex dynamics in soft condensed matter, that are usually characterized by high degrees of confinement, significant molecular crowding, active propulsion, and heterogeneous material and dynamical properties. Our study explores how these combined factors, constituting a highly complex medium away from simple equilibrium scenarios, influence the emergent dynamics and mechanical responses of the model systems and whether any universal behavior exist underlying particle motion for these cases.}

\subsection{3.1 Dynamic heterogeneity in supercooled liquid} 
{When a liquid is supercooled close to its glass transition, dynamics become increasingly heterogeneous \cite{binder2011glassy, berthier2011dynamical, berthier2011theoretical}. Particle motion slows down, and mobility varies significantly across regions, leading to spatially heterogeneous relaxation behavior. Fast-moving particles coexist with constrained ones, forming dynamic domains with distinct mobility scales. This coexistence reflects a breakdown of uniform relaxation and is often attributed to transient structural constraints or local free energy barriers. The onset and persistence of this heterogeneity depend on the strength of thermal fluctuations and their ability to facilitate cage escape, balancing local confinement with entropic driving, to achieve diffusion at the cost of free energy\cite{wang2012brownian, karmakar2014growing, vaibhav2024entropic}.} {During the coarse of its motion, particles traverse complex energy landscapes, and Fickian diffusion emerges only at timescales much longer than the typical observation intervals—where mean-squared displacement regains linearity.}. 

In Fig.~\ref{fig1}(a), we show the spatial maps of particle displacements {(Equation~(\ref{eq:disp_map}))}, when ambient temperature is close to its glass transition. {The map reveals coexisting regions of high and low particle mobility, indicative of spatial heterogeneity.} This heterogeneity becomes more prominent at lower temperatures. The typical residence time to be {\it locked-in} within the cage is roughly estimated by the plateau length in mean-squared displacements {(Equation~(\ref{eq:msd}))}. The long time diffusion sets in when the mean-squared displacement becomes proportional to time {[Fig.~\ref{fig1}(b)]}. At low temperatures, the displacement distributions {(Equation~(\ref{eq:disp_dist}))} show exponential tails, even when mean-squared displacement grows linearly in time, persisting for very long times [Fig.~\ref{fig1}(d)], while they revert back to Gaussian at a much smaller interval at high temperatures [Fig.~\ref{fig1}(c)] \cite{berthier2011dynamical}. The conventional measurement of the persistence of this characteristic timescale (over which the displacement distribution remains non-Gaussian) is typically quantified by the kurtosis of the displacements {(Equation~(\ref{eq:alpha2}))}, which shows {a higher peak} at lower temperatures and persists strongly non-Gaussian for much longer spans [Fig.~\ref{fig1}(e)]. For the same set of the displacements at a given temperature, we show that Negentropy {(Equation~(\ref{eq:defng_condensed}))} \cite{shell2016coarse, vaibhav2024entropic} provides a more robust measurement of the lifetime of the persisting heterogeneity [Fig.~\ref{fig1}(f)].

{The relaxation of local structures and their statistical signatures are intrinsically connected. Exponential tails in displacement distributions reflect low probable fast particles making large, cage-breaking jumps, while the trapped majority contributes to the central peak. This temporal heterogeneity evolves non-monotonically defining its scale of persistence: non-Gaussianity peaks at the cage lifetime, marking maximal differentiation before collective relaxation restores Gaussian behavior. The conventional non-Gaussian parameter captures the information centrally to this Gaussianity, while negentropy, in contrast, considers deviations from all orders, offers a more robust quantification of non-Gaussian dynamics and the persistence of heterogeneity.}

\subsection{3.2 Amorphous plasticity in squishy systems}
\begin{figure}[t!]
    \centering
    \includegraphics[width=1.0 \linewidth]{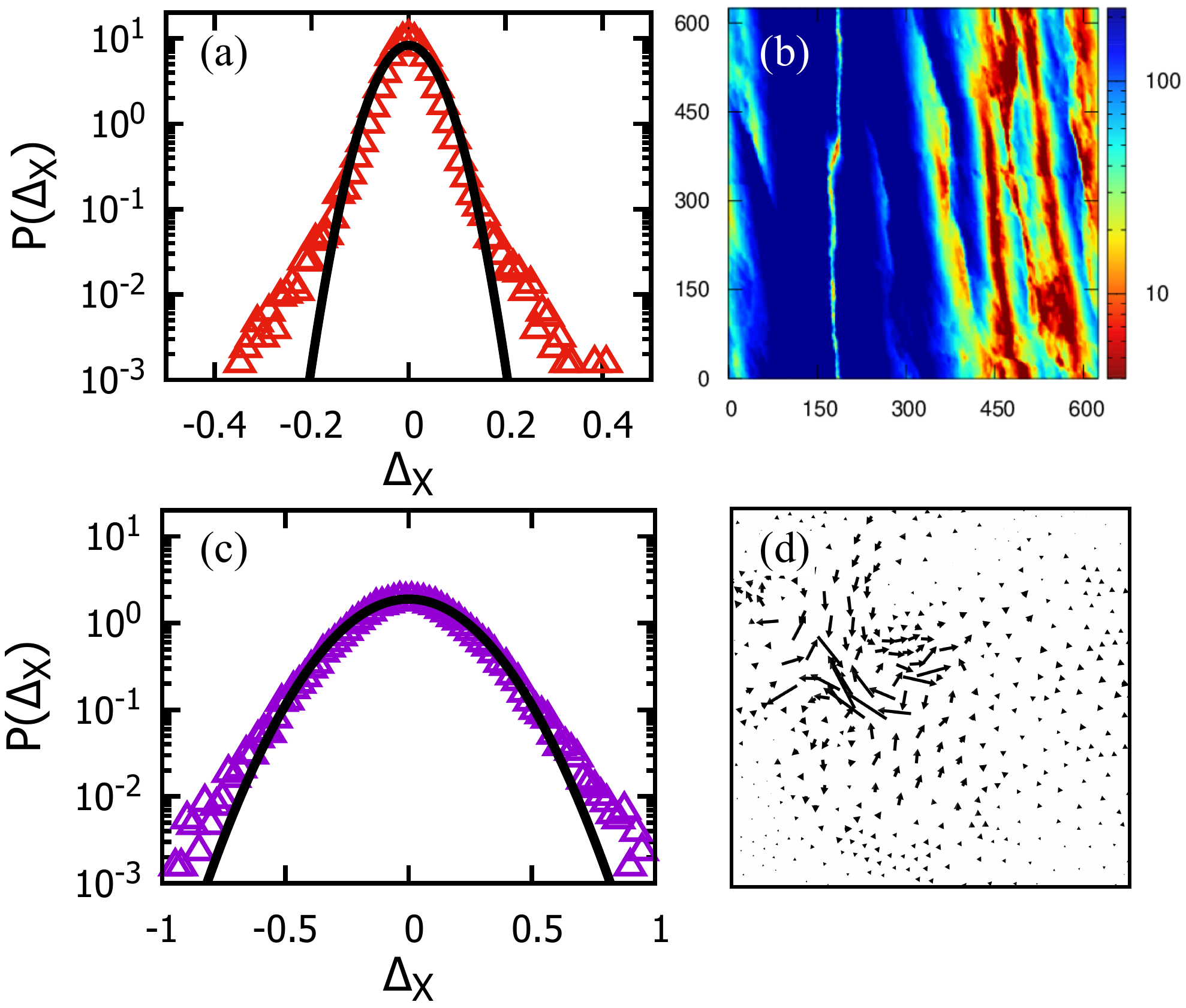}
    \caption{Plastic rearrangements as a precursor to catastrophic failure in yield stress materials : (a) Distribution of instantaneous particle displacements ($P(\Delta_{X})$) temporally close to critical failure, for an imposed shear-stress $\Sigma=0.85$ (in open symbols). The solid line shows the nearest Gaussian distribution respective to $P(\Delta_{X})$. (b) Spatial map of coarse-grained mean-squared-displacements shows regions of kinetically active sites and presence of percolating back-bone. (c) Instantaneous displacement distributions in presence of steady flow for $\Sigma (= 0.85) > \Sigma_{Y} ( \approx 0.81)$, where $\Sigma_{Y}$ is the yield strength. The solid line shows the respective nearest Gaussian. (d) The corresponding plastic rearrangements are shown by local displacement map, which are non-affine in nature. The data has been obtained for a system of size $N=102400$ in two dimensions using Model 2.}
    \label{fig2}
\end{figure}

\begin{figure*}[t]
    \centering
    \includegraphics[width=0.8\linewidth]{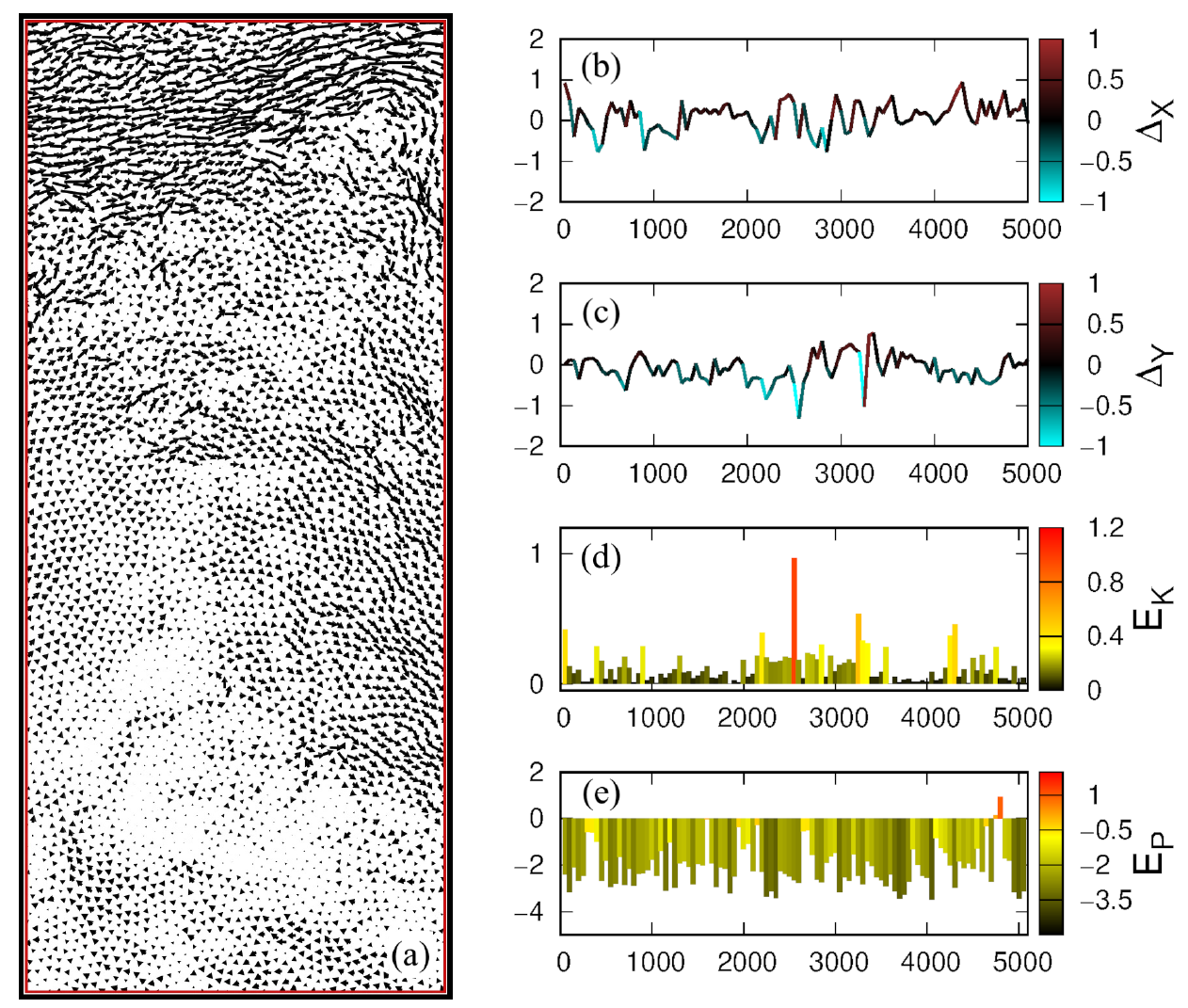}
    \caption{Dynamic Facilitation in Active Glassy Liquid: (a) Coexistence of slow and fast regions. Spatial maps of particle displacements are shown in arrows. It shows that regions of collective flow and arrested regions coexist together within the same system. Time series of a typical single particle instantaneous displacements in x (b) and y-directions (c) are shown. Associated time series of (d) kinetic energy and (e) potential energy for the specified particle shows the intermittent nature of the dynamics and persistent heterogeneity and occurrence of extreme events in the form of kinetic bursts. The underlying extreme statistics is clearly seen with unusual spikes in the time-series of displacement and kinetic energies, falling on a distribution which is non-Gaussian (data not shown). The data was simulated using Model 3 for a system size $N=100000$ in two dimensions with $f=3$. The system has a threshold for unjamming transition at $f_{C}\approx1.67$ for $N=1000$ as reported in Ref. \cite{mandal2020extreme, dutta2025activity}.}
    \label{fig3}
\end{figure*}

Squishy materials exhibit a unique mechanical response to applied stress, remaining rigid below a critical threshold before undergoing significant plastic deformation and yield beyond it. This phenomenon has garnered significant attention due to its relevance across diverse applications, from materials processing and biomedical engineering to food-rheology and consumer products \cite{nicolas2018deformation, ozawa2018random, berthier2025yielding}. We investigate creep using molecular dynamics simulations, wherein amorphous solids are kept under applied constant stress, in complete absence of thermal fluctuations, to test whether non-thermal and non-affine plastic fluctuations can yield a material \cite{cabriolu2019precursors, liu2021elastoplastic, dutta2023creep}.

In Fig.~\ref{fig2}, we show that extreme particle displacements play a critical role in onset of flow {within} a system that shows emergent shear{-induced deformations} under constant imposed stress. During the yielding, instantaneous particle displacements are non-Gaussian, and the distribution shows {\it fat tail} due to the presence of large displacements beyond normality. This is shown in Fig.~\ref{fig2}(a). The system shows {onset of flows when imposed stress crosses a critical yield stress. The spatial regions with
large activity is prominently seen [Fig.~\ref{fig2}(b)]. Nearby the yield threshold, we also observe a largely diverse spatial maps of the displacements }[computed using {Equation~(\ref{eq:local_msd})}]. The displacement in post-yield sites also show non-Gaussian behaviour [Fig.~\ref{fig2}(c)]. This onset of flow in these systems is a consequence of {\it non-affine} plastic rearrangements that cascade through an avalanche of events \cite{ozawa2018random, cabriolu2019precursors}, systematically failing the system at macro-scale. The local structure of the plastic events can be seen in Fig.~\ref{fig2}(d). These local rearrangements are usual consequence of {\it Eshelby}-like local stress redistribution with quadrupolar symmetry \cite{dasgupta2012microscopic, nicolas2018deformation, liu2021elastoplastic}. This suggests that the heterogeneities in particle dynamics are caused by extreme fluctuations, populating particles from one minimum to another. The surprise is that these extreme displacements persist even in the steady states in the fluidized state.

{The persistence of these extreme, non-Gaussian displacements even in the steady-state flow regime suggests a deeper connection to the dynamics of the thermal super-cooled liquids, where structural relaxation is governed by intermittent, spatially heterogeneous rearrangements. In our athermal creep simulations, the onset of flow via non-affine plastic events and Eshelby-like stress redistribution mirrors the activated hopping processes seen in the passive liquids near the glass transition. The fat-tailed displacement distributions and localized mobility suggests a facilitation-driven mechanism for spatio-temporal heterogeneity under imposed stress.}

\subsection{3.3 Self-regulated {active} flow and dynamic facilitation}
In flowing active materials, extreme displacements play a crucial role in controlling the lifespan of microscopic heterogeneity and the degree of coordination. Understanding such complex flow patterns are thus, extremely useful to reverse-engineer materials with tunable dynamical and topological properties \cite{gompper20202020, debets2023glassy}. For instance, the emergence of collective dynamics in these systems are shown to be facilitated by local particle rearrangements that are irreversible and breaks detailed balance \cite{marchetti2013hydrodynamics}. These extremely unusual, yet prevalent moves drive particles to hop intermittently over the free energy barriers within the spatiotemporally evolving landscape, even in the complete absence of the thermal fluctuations. 

In Fig.~\ref{fig3}(a), we show autonomous collective motion in athermal dense systems of self-propelled particles. The system experiences strong dynamic heterogeneity as parts of the systems flow much slower than its neighboring regions. We show the timeseries of displacements {(Equation~(\ref{eq:disp_series}))} for one of the tracked particles within a self-regulated domains (chosen randomly) in two different spatial dimensions in Figs.~\ref{fig3}(b) and (c). The associated kinetic and potential energy timeseries are shown in Figs.~\ref{fig3}(d) and 3(e). They all suggest that the dynamics is intermittent and sudden facilitated movements are accompanied by large kinetic bursts in the system. The changes could be prominently seen in Figs.~\ref{fig3}(d-e). However, these moves are extremely low probable and suggest that these activities provide meaningful escape-routes between the temporally evolving minima within the rough landscape, even when they are low probable and intermittent.

{ The spatio-temporal heterogeneity observed in flowing active material as modeled here reveals a striking interplay between localized rearrangements and emergent global coordination. Intermittent bursts of kinetic energy, as shown in Fig.~\ref{fig3}, are due to facilitated displacements that allow particles to traverse together between evolving minima. These transitions are not uniformly distributed in space or time; instead, they emerge as dynamically heterogeneous zones where motion undergoes localised trapping and spontaneous activation. Such behavior reflects a landscape-driven facilitation mechanism, where the system self-organizes into regions of high and low activity, reminiscent of glassy relaxation and activated dynamics in dense active liquids where rare, extreme events regulate the lifespan of microscopic heterogeneity and sustain flow—highlighting a self-organized pathway to yielding that connects mechanics of glassy systems and physics of fluids.}

\section{4. Discussion and Conclusion}
{In summary, we explore the emergence of spatio-temporal heterogeneity in a variety of amorphous systems subjected to distinct physical conditions. These systems—ranging from passive to stressed and actively driven glassy states—are influenced by either thermal or athermal fluctuations, and are simulated at extreme densities using molecular dynamics. By analyzing single-particle displacement statistics, we uncover how particle motion in dense metastable liquids universally show persistent deviations from typical Fickian diffusion.}

{Early molecular dynamics simulations revealed non-Gaussian dynamics in simple liquids, including hard-sphere fluids exhibiting power-law decay of velocity correlations due to microscopic vortex formation and collective diffusion \cite{rahman1962theory, rahman1964correlations, alder1970decay}. In the context of glassy systems, model studies further linked non-Gaussian dynamics to anomalous, discrete jump statistics in the vanishing-diffusion limit \cite{odagaki1991gaussian, bouchaud1990anomalous}. For more realistic systems, such heterogeneity has been attributed to the rugged nature of the potential energy landscape \cite{sastry1998signatures}. This notion was later generalized to systems approaching jamming or the glass transition, where non-Gaussian dynamics appear to be universal \cite{chaudhuri2007universal}. Even in driven systems, such behavior remains prevalent \cite{vasisht2018rate, dutta2018transient}.}

{Deviations from Gaussianity  usually quantified by {\it non-Gaussian parameter} that captures the excess of the fourth moment relative to the square of the second moment \cite{berthier2011dynamical}. Yet, despite being successfully applied over decades, this simple probe was recently shown to be inadequate when the heterogeneity is strong and intermittent particle dynamics control the amplitude and persistence of heterogeneity \cite{vaibhav2024entropic}, because moment-based metrics, typically truncated at low orders, are lesser sensitive to rare fluctuations that usually appears at the tails. In contrast, the information-theoretic measure Negentropy quantifies non-Gaussianity based on all order moments of the data, in terms of statistical differences between the non-Gaussian distribution relative to its nearest Gaussian distribution \cite{vaibhav2024entropic, ivan2012measure}. In the context of quantifying the persistence of dynamic heterogeneity, therefore, incorporation of rare and sensitive fluctuations is important and informative approaches naturally outperform the moment based ratios \cite{vaibhav2024entropic, dandekar2020non}. }

{Across passive, sheared, and active systems, slow relaxation and dynamic heterogeneity arise from particle dynamics within complex topological landscapes—whether energetic, motility-driven, or flow-induced \cite{vaibhav2024entropic, dutta2023creep, mandal2020extreme, dutta2025activity}. Near the glass or jamming transition, particles become intermittently trapped in metastable basins, requiring rare fluctuations to overcome activation barriers \cite{berthier2011dynamical, karmakar2014growing, vaibhav2025experimental}. These activated processes govern the persistence and life time of such dynamic heterogeneity. In sheared materials and dense cellular assemblies, plastic rearrangements reflect transitions between topologically distinct configurations, often triggered by stress or ambient fluctuations. By breaking time-reversal symmetry, active agents like motile cells and microswimmers adapt their propulsion strategies to navigate complex, spatially varying motility landscapes \cite{Lowen2025-if, Kant2025-kb}. In both biological and synthetic systems, these interplay of activity, crowding, and mechanical forcing often generates turbulent-like flows—characterized by vorticity, energy cascades, and emergent mesoscale structures, governing the topological transitions \cite{berthier2011dynamical, karmakar2014growing, vaibhav2025experimental, mukherjee2023intermittency, Kant2025-kb}.}

{Traditional statistical approaches often fail to predict the full extent of such spatio-temporal heterogeneities, especially close to criticality, where extreme events marginally separate states of flow and dynamic arrest. In this context, machine intelligence offers a transformative framework for characterizing and decoding such complexity: by reverse-engineering displacement spectra, identifying latent mobility patterns, and mapping the topology of metastable basins. Data-informed approaches can also identify optimal navigation strategies, reveal hidden structures in particle dynamics and facilitate predictive modeling of non-equilibrium transitions \cite{McDermott2019-af, Lowen2025-if, Dulaney2021-qy}.} {The present work, by decoding the physics of non-Gaussianity under out-of-equilibrium scenario, guides further explorations at the interface of physics and machine intelligence. It not only deepens scope to address realistic complex problems using hybrid technologies, but also opens explainable pathways for inverse-designing adaptive materials.} 

\medskip
\textbf{Acknowledgements} \par
VV acknowledges the computational resource provided via the project ``TPLAMECH'' on INDACO platform at the HPC facility of Università degli Studi di Milano. TD is supported by World Premier International Research Center Initiative (WPI), MEXT, Japan. SD acknowledge HPC facilities of ICTS-TIFR where the computations have been performed under the Department of Atomic Energy (GOI) project No RTI 4001. Part of this work was carried out during SD's visit to Indian Statistical Institute, Kolkata and ICTS-TIFR.

\medskip
\textbf{Research Contributions} \par
V.V.: Data Curation and Analysis (Part); Visualizations, Discussion, Review, and Editing. T.D.: Discussion, Review, and Writing (Part). S.D.: Conceptualization; Writing; Review and Editing; Data Curation and Analysis (Part), Interpretation, and Analysis; Visualizations, Presentation, and Project Coordination.

\medskip
\textbf{Conflict of Interest}\par
The authors declare no conflict of interest.

\medskip
\textbf{Data Availability Statement}\par
The data that support the findings of this study are available from the corresponding author upon reasonable request.

\bibliography{references.bib}

\end{document}